\documentclass{aastex631}
\graphicspath{{graphics/}}
\newcommand\myplot[2]{\makebox[\textwidth][c]{\includegraphics[width=#1\textwidth]{#2}}}

\shorttitle{Characterization of the SN Spectral-Temporal Energy Distribution}
\shortauthors{Bengyat \& Gal-Yam}

\submitjournal{\apj}
\received{21 Feb 2022}
\revised{24 Mar 2022}
\accepted{25 Mar 2022}

\begin{document}

\title{Characterization of Supernovae Based on the Spectral-Temporal Energy Distribution: Possible two SN Ib Subtypes}

\author[0000-0001-5547-9176]{Ofek Bengyat}
\affiliation{Department of Particle Physics and Astrophysics, Weizmann Institute of Science, 76100 Rehovot, Israel}
\affiliation{Faculty of Physics, University of Vienna, Boltzmanngasse 5, 1090 Vienna, Austria}

\author[0000-0002-3653-5598]{Avishay Gal-Yam}
\affiliation{Department of Particle Physics and Astrophysics, Weizmann Institute of Science, 76100 Rehovot, Israel}

\begin{abstract}
    A quantitative data-driven comparison among supernovae (SNe) based on their spectral time series combined with multi-band photometry is presented. We use an unsupervised Random Forest algorithm as a metric on a set of 82 well-documented SNe representing all the main spectroscopic types, in order to embed these in an abstract metric space reflecting shared correlations between the objects. We visualize the resulting metric space in 3D, revealing strong agreement with the current spectroscopic classification scheme. The embedding splits Type Ib supernovae into two groups, with one subgroup exhibiting broader, less prominent, higher-velocity lines than the other, possibly suggesting a new SN Ib subclass is required. The method could be to classify newly discovered SNe according to their distance from known event groups, or ultimately to devise a new, spectral-temporal classification scheme. Such an embedding could also depend on hidden parameters which may perhaps be physically interpretable.
\end{abstract}

\section{Introduction}
The existing classification scheme of supernovae \citep[SNe,][]{filippenko_1997, Gal-Yam2017} has been successful in sorting out the majority of objects discovered, in a way that facilitates their study or practical use, e.g. as standard candles \citep{1998AJ....116.1009R, Perlmutter_1999}. It does that by relying mainly on their spectra.

The current classification scheme divides SNe to Type I and Type II, comprised of SNe lacking H-lines and containing them, respectively. The reason for this is mostly historical \citep{minkowski1941}, and modern understanding distinguishes between two other categories based on the progenitor star and the underlying physical explosion process. The first class is thermonuclear SNe, including mostly members of the spectroscopic Type Ia group, that were found \citep{2011Natur.480..344N} to be a result of a white dwarf (WDs) progenitor, exploding  in a manner still actively researched \citep{Ia_research1,Ia_research2}. Spectroscopically, normal Type Ia SNe are identifiable by the strong {\ion{Si}{2}~6355\AA} line seen in peak spectra; though this is less clear in some sub-types. The rarer group of Ca-rich Type I SNe has also been associated with long-lived, WD progenitors \citep{Perets2010,De2020}. The second category is core-collapse SNe, originating from massive stars whose self-gravity ceases to be supported by nuclear fusion in their core. These are further divided into the H-rich Types II and IIn, the latter distinguished by their narrow Balmer emission lines ascribed to circumstellar material (CSM), and the stripped-envelope SN Types, which are the explosive deaths of massive stars which have already lost their outer hydrogen shell prior to explosion. These include Types Ib and Ic, the latter differing by the lack of He in peak spectra (in particular, the {5876\AA}, {6678\AA}, and {7065\AA} lines). Type IIb SNe exhibit hydrogen lines only in early spectra, which later disappear, indicating a much thinner layer of H in their progenitor star than those of normal Type II SNe. He-poor stripped-envelope SNe with broad spectral lines belong to Type Ic-BL, corresponding to explosions with larger kinetic energy per unit mass, and hence higher ejecta velocity -- up to about threefold compared to the $\sim 10^4 \,\mathrm{km/s}$ of normal Type Ic SNe. While these are the main, most frequently observed Types, others exist, including stripped-envelop SNe with CSM interaction of Types Ibn \citep[e.g.,][]{Pastorello2008} and Icn \citep{Gal-Yam2021}, superluminous SNe \citep[e.g.,][]{Gal-Yam2019}, as well as numerous subdivisions of the Types mentioned here. A recent review about the current classification scheme of SNe could be found in \citet{Gal-Yam2017}.

In practice, an experienced individual could examine a spectrum of some SN, mainly by looking at emission or absorption lines, determine its type and enter it to some database. The problem of subjectivity that could possibly emerge in this human classifier scenario has been treated by developing tools such as \texttt{Superfit} \citep{superfit} or \texttt{SNID} \citep{snid}, both relying on a template bank of SN spectra to which a query spectrum is compared. Considerations such as redshift, reddening through Milky Way or host dust, photometric calibration of the spectrum and contamination by host lines are present in these methods as fit parameters or have their effects diminished by smoothening or continuum estimation and subtraction. These methods, as opposed to human classification, standardize the treatment of different SNe such that different reddening or contamination conditions have less impact on the classification, and they are also more objective, being independent of the user, his attention to details or other forms of human error. The use of such automatic tools also streamlines the task of classification of large amounts of data. Another quantitative approach to classification, of Type I SNe, also relying on peak spectra, was presented by \citet{sun2017}. This classification uses the estimation of the depths of a line around {6150\AA} attributable to \ion{Si}{2} or H, depending on the SN, and of {\ion{O}{1}~7774\AA} and achieves significant separation between the main Type I SN subtypes (Ia, Ib and Ic), as well as between different Ia subtypes.

However, one drawback in the presently available schemes and methods which still awaits remedy is that they do not rely on the entirety of information available for a SN, but on a single spectrum of it, i.e. from a single epoch. This could also lead to ambiguity, this time not a subjective, user-dependent one, but the mere fact that the current scheme could assign different types to the same object when faced with different spectra of it. An example is Type IIb, which includes core-collapse SNe exhibiting H in early spectra but not in later ones. \citet{phys_mot} have approached the study of this type-continuum by distinguishing between 4 sub-types ranging from Ib to IIb based on the $\mathrm{H}\alpha$ emission and absorption lines. They also treated the continuum between the He-poor stripped core-collapse SNe Types Ic and Ic-BL, the latter being defined by significantly higher explosion velocities than the former, evident in the breadth of spectral lines. This difference in velocities is, just as the Ib--IIb case, not only a continuous parameter discretized into two groups, but also a time-dependent property, as they show that it changes throughout the lifetime of the transient. They do this by examining in practice the related (though only weakly correlated) property of feature count, and divide the continuum into 5 subtypes depending on the average of the feature count over a certain time interval. Examples for He-poor stripped core-collapse events of uncertain classification are SN2004aw \citep{04aw} and SN2016coi \citep{2018MNRAS.473.3776K}. Additionally, a certain confusion in the classification of stripped core-collapse SNe as Ib or Ic exists. One reason for it is the presence of He, observed in some events initially classified as Type Ic (e.g., SN2007gr, SN2009jf). This has been perceived as a flaw in the current classification scheme, rather than an incomplete physical understanding of the objects \citep{on_Ibc,09jf_07gr,snid}.

There exists then a certain need for a classification scheme both quantitative and taking into account the entire \emph{spectral-temporal} energy distribution of a SN (i.e., the spectral flux $f(t,\lambda)$ along both wavelength and time) in order to provide a more definite answer about the degree of similarity between, and thus the proper taxonomic sorting of SNe. Such a method could perhaps further unveil some structure in the space of SNe, possibly in the form of a few hidden parameters controlling their properties. Those might, ideally, be later ascribed to physical properties through correlation analysis.

\subsection{Machine Learning Methods in Astronomy}
Various methods of data analysis have been used in astronomy, the emphasis being put in recent years on machine learning methods. A few examples for uses in the context of SN classification are \citet{sasdelli+2014_Ia_metric,sasdelli+2016_Ia_autoencoder,Williamson_2019}. \citet{fremling2021sniascore} presents a binary deep-classifier named \texttt{SNIascore}, which succeeds in classifying SNe as Type Ia with very low error rates. See also \citet{baron2019machine} for a useful review on machine learning methods in astronomy, including those used in this work. The question of quantitative SN classification could be approached by means of some machine learning algorithm which would be trained on feature vectors describing a set of SNe. This training set should then ideally consist of well-sampled SNe, whose classification and comparison to other objects is well-understood. After training, the algorithm should be able to provide new insight about this taxonomy of and similarity between objects when input SNe from or outside of the training set. The method we use in this work is an unsupervised Random Forest (RF) serving as a metric \citep{shi_horvath_2006}, which has been found viable in defining a distance between spectra \citep{baron_rfalg,reis2018a}. Section~\ref{subsec:rf} describes the algorithm and its application.

\section{Objective} 

In this work, we present a method to characterize, or classify, SNe based on their spectral-temporal energy distribution by means of a data-driven embedding in an abstract metric space. We compare the spectroscopic types in the current scheme -- as determined by a human expert after examining the spectral sequence available for the SN in question -- with the resulting embedding.

Other than this comparison, we examine possible Type-continua or new subtypes which may arise from the embedding.

\section{Method}
\subsection{Estimating the Spectral-Temporal Energy Distribution of a SN}\label{subsec:pycoco}

\begin{figure*}
\myplot{1}{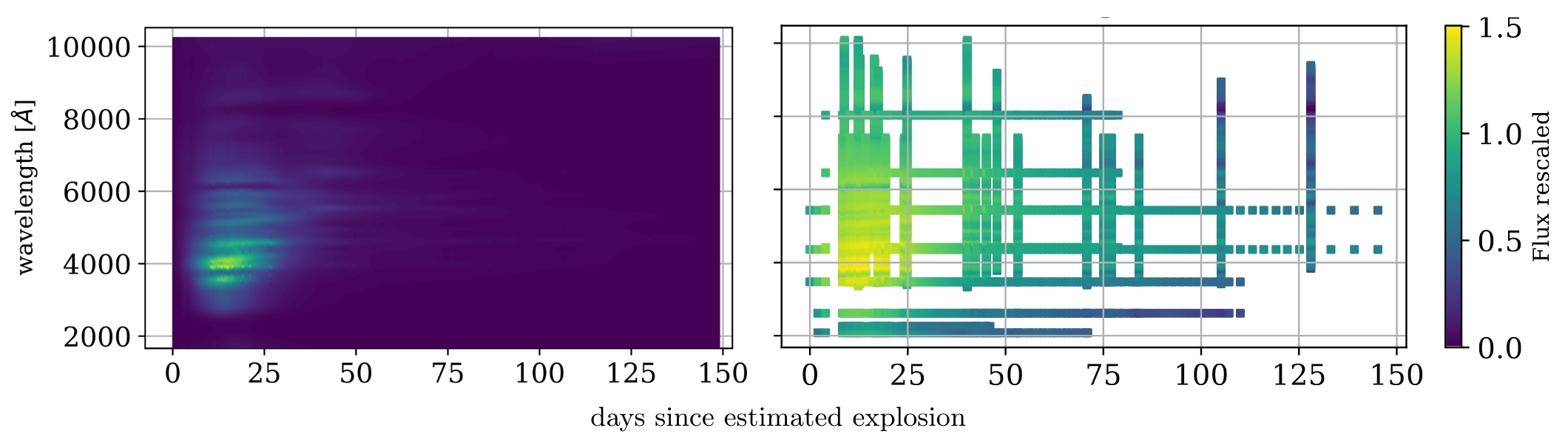}
    \caption{An example of the output (left) and input (right) of \texttt{PyCoCo} for SN2007af. The interpolated, calibrated and dereddened rest-frame flux function $f(t, \lambda)$ can be seen on the left, and the input spectral and photometric data is on the right. Note that the colors on the right panel differ from those on the left as the input spectra and photometry are not yet calibrated or dereddened.}
    \label{fig:pycoco_example}
\end{figure*}

The raw data of each SN, i.e. its spectra and multi-band photometry, are processed with \texttt{PyCoCo}, a program for template creation of spectral energy distribution of SNe developed by \citet[][hereafter V19]{Vincenzi_2019}. It uses Gaussian Process Regression to interpolate the flux in the intervals between the input spectra, with the photometry data used both for flux calibration of the spectra and as data points. \texttt{PyCoCo} also corrects the flux for Milky Way extinction and host extinction\footnote{Host uncorrected output is also produced, though in this work we only use the host corrected versions.} values the user inputs, and removes tellurics and host lines, as well as transforming the data into the rest frame of the SN. The final output from \texttt{PyCoCo} is the (normalized) rest-frame spectral energy distribution $f(t, \lambda)$ of the input SN on some grid of time and wavelength.

We keep the power $n$ of the light-curve rising fit in \texttt{PyCoCo}, $f \propto (t-t_0)^n$, (V19, Eq.~2) as a free parameter for all the SNe we apply it on, and we also choose to not extend and oversample the light-curves at late times, as our analysis concentrates on relatively early times (see next section). As the majority of \texttt{PyCoCo} outputs used in this work were generated by V19 (see Section~\ref{sec:data}), it should be mentioned that they only left $n$ as a free parameter in cases where the photometric data shortly after explosion was good enough to allow a reasonable fit. When not, $n$ was fixed to a value, which depends on whether the SN is H-rich (Types II and IIn) or not (the rest of the types), as indicated by the user. This is the only place where input regarding classification is used in \texttt{PyCoCo}. It should also be noted that a different light curve rise model was used for the data sets of Type IIb  SN1993J, SN2011dh, SN2011fu, and SN2013df generated by V19, in order to take into account their double early peaks. This is also true for SN2006aj, but in this case its double peak is likely a result of its unusually early detection, rather then an actual peculiarity \citep{06aj}, so we omit the first peak data for this object. Readers are referred to Section~2.1 in V19 for details about the light-curve fits in \texttt{PyCoCo}.

The code used in this section is available on GitHub\footnote{\url{https://github.com/ofek-b/PyCoCo_templates}}. See Figure~\ref{fig:pycoco_example} for an example of the process.

\subsection{Preparation of Feature Vectors}\label{subsec:prep}
We use an algorithm from \citet{spectres}\footnote{\url{https://github.com/ACCarnall/spectres}} to rebin in wavelength and perform linear interpolation in time to obtain the fluxes on a uniform grid $T \times \Lambda$ over all SNe, where
\begin{center}
    $T = $ 0 d to 50 d at 1 d intervals\\
    $\Lambda = $ 4000 {\AA} to 8000 {\AA} at 40 {\AA} intervals.
\end{center}
Times are counted since the estimated explosion of each template. We choose this wavelength interval $\Lambda$ as it is covered available photometric data for almost all SNe in our data set, making the output of \texttt{PyCoCo} more credible. The time interval $T$ is chosen so it includes times where most SNe are not too faint.

As others have done (e.g., \citealt{sasdelli+2014_Ia_metric}, or less directly in \citealt{snid}), we use the wavelength derivative of the logarithm of the flux of every SN,
$$
\tilde{f}(t,\lambda) \equiv \frac{\mathrm{d}}{\mathrm{d}\lambda}\mathrm{log}f(t,\lambda),
$$
instead of the flux itself, in order to emphasize spectral lines and diminish the effect of the unaccounted for reddening. The data is then sorted in an $N$-row matrix $X$, where each row $X_i$ corresponds to the $i$-th SN in our data set:
$$
X_i = (\tilde{f}(t_j, \lambda_j))_{(t_j, \lambda_j) \in T \times \Lambda}.
$$

In order to account for some missing data points we then use the Expectation Maximization Principle Component Analysis (EMPCA) algorithm presented in \citet{empca}\footnote{\url{https://github.com/sbailey/empca}} on our feature matrix $X$ to produce a principal components matrix, $P_C$, which is then immediately transformed back to the feature space to produce $X_C$, now without missing data. The number of principal components we choose is $C = 50$, which preserves 93\% of the variance.

\subsection{Unsupervised Random Forest as a Metric}\label{subsec:rf}

\begin{figure*}
    \myplot{0.7}{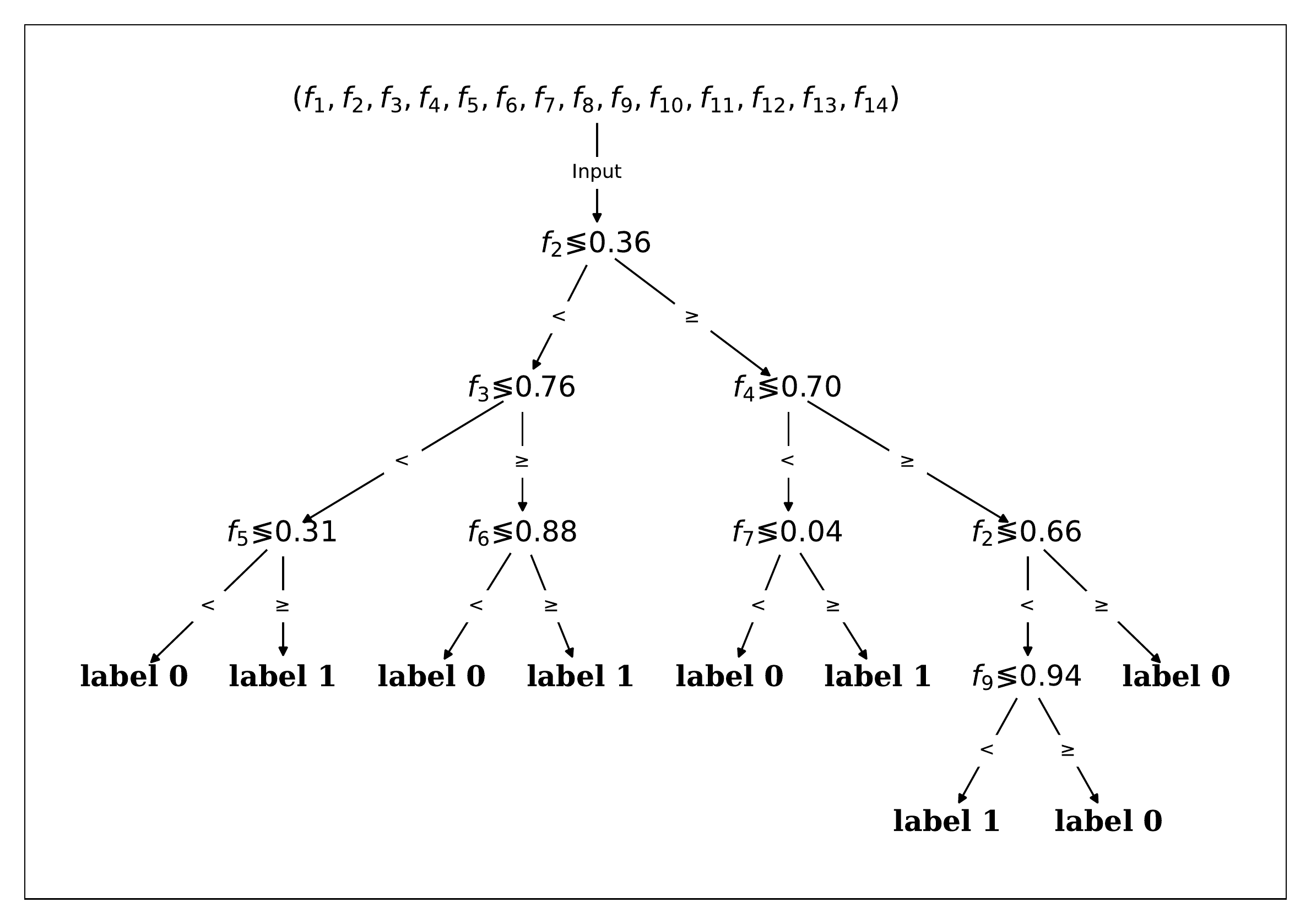}
    \caption{An illustration of a trained binary Decision Tree, multiple different instances of which comprise an RF. The input object propagates according to its features \textsf{f}$_i$.}
    \label{fig:tree}
\end{figure*}

We now train an unsupervised RF on the data in $X_C$ and use it as a metric on the set of SNe as described by the same matrix $X_C$. This section describes this process.

RF is originally a supervised machine learning classifier \citep{ho95}. A brief description of it follows, regarding only binary classification as relevant for this work. First, we describe binary Decision Trees. A Decision Tree is a supervised classifier by itself, structured as a tree graph -- a directed graph with a single start node, several terminal nodes and several "generations" of intermediate nodes in between, where every intermediate node has exactly one arrow pointing to it from its parent and exactly two arrows which it points to its children. When an object is queried into the Decision Tree, each non-terminal node $k$ hands it over to one of its two child nodes, chosen according to the truth value of a condition of the form $f_{i_k} < c_k$, where $f_{i_k}$ is the value of the $i_k$-th feature of the Tree input, and $c_k$ is a real number. The training process consists of the simultaneous input of all objects in the training set. Before propagating a set of objects which has arrived to any node $k$, the training algorithm chooses the parameters $i_k$ and $c_k$ providing the best separation between the known labels for the objects. The definition of goodness of separation is a matter of choice (a so-called hyperparameter). When all of the objects which arrived in a node are of the same label, this node is set as a terminal node and is assigned the label. Training ceases when all objects arrived at terminal nodes. Classifying a queried object occurs by propagating it through the trained Decision Tree and returning the label corresponding to the terminal node it reaches. Note that Decision Trees and their training process are deterministic. See Figure~\ref{fig:tree} for an illustration and the classification scheme for Type~I~SNe presented in \citet[][Section~4.2]{sun2017} for a working example.

An RF consists of multiple Decision Trees. Every Tree is (i) only trained on a random subset of the training set, and (ii) only allowed to use a random subset of the features. The class assigned to a queried object is determined by a majority vote of all the Decision Trees in the Forest. An RF is thus an ensemble learning method, aggregating the results of the Trees to achieve better accuracy and robustness. In order to use an RF as a metric, one performs the following \emph{unsupervised} training process. The training data set in question is used to create a synthetic data set of the same size. The samples in the synthetic data set have the same number of features as in the original data set and the same marginal distributions for each feature, but zero correlations between features. Supervised training follows, as described above, using the real and synthetic sets, where the real set is assigned the label \texttt{real} and the synthetic set is assigned the label \texttt{synth}. Note that these labels have nothing to do with SN types or with the data itself in any way, and they merely differentiate between the real, correlation exhibiting data, and the synthetic, uncorrelated data. After training, the similarity returned for two queried objects is the fraction of Trees in the Forest for which the two objects end up in the same terminal node, out of only the Trees which assign label \texttt{real} for both objects (i.e., the correct label, as one tries to find the similarity measure between two real objects). The metric, distance or dissimilarity (used interchangeably in this work) is obtained by subtracting the similarity from unity.

This similarity measure thus learns the most prominent correlations in the training data, and measures to which extent do two given objects share such correlations. One such correlation could be, for instance, the presence of all three strongest He lines in a spectrum, as opposed to only some of them.

The algorithm we use was written by \citet{baron_rfalg}\footnote{\url{https://github.com/dalya/WeirdestGalaxies}}. The number of Trees used is 2000, and the separability criterion is the Gini Impurity. The output of our analysis is the distance matrix for our SN data set, with entries between 0 and 1.

The code used in Sections~\ref{subsec:prep} and \ref{subsec:rf}, as well as in the following visualization of the results is available on GitHub\footnote{\url{https://github.com/ofek-b/spectra_in_time}}.

\section{Data}\label{sec:data}

Alongside the description of the  \texttt{PyCoCo} code, V19 also includes the code output for 67 core-collapse SNe, listed in Table~2 of their publication. Our data set is built upon those SNe, with some exclusions and type changes of SNe listed in Table~\ref{tab:vincmod}. We extend this data set by adding the SNe listed in Table~\ref{tab:mydata}. The final list of 82 SNe used in this work is given in Table~\ref{tab:alldata}, along with the Data Quality Index (DQI) described next. This data set will be made available upon request.

Two SN Types missing from the V19 set, which we also did not add, are Ca-rich SNe and H-poor superluminous SNe (SLSN-I). In the case of Ca-rich SNe, the reason is the poor data sets available for those objects. In the case of SLSN-I, peak light often occurs late, and outside our temporal window, such that our selected time grid would miss important information about those objects, the same holds also for SN1987A-like Type II SNe. One could envision including these groups once more data are available, and including a temporal renormalization for SLSNe-I and 87A-like events.

\begin{table*}
\centering
\caption{The SNe from V19 which were excluded from the training set in this work or whose assigned spectroscopic type was changed after examining their spectral series.\label{tab:vincmod}}
\begin{tabular}{lll}
\tableline
Name & Type in V19 & Comments\\
\tableline
\textit{Excluded:}\\
SN1987A & 87A-like & long rise time with peak outside our temporal range\\
SN2005bf & Ib & very anomalous, double-peaked light curve \citep{05bf}\\
SN2008D & Ib & highly extinguished \citep{rabinak_waxman}; unique early time data\\
SN2011bm & Ic & very wide light curve \citep{11bm}\\
SN2016bkv & II & faint, unusually long light curve, weaker lines \citep{16bkv}\\
SN2008in & II & ${\rm DQI} < 0.8$ \\
SN2009dd & II & ${\rm DQI} < 0.8$ \\
SN2013fs & II & added again manually for the analysis in Section~\ref{sec:displ}. appears on Table~\ref{tab:mydata}. \\
SN2008aq & IIb & ${\rm DQI} < 0.8$ \\
SN2009ip & IIn & peculiar object, may not be a SN \citep{09ip}\\
SN2011ht & IIn & ${\rm DQI} < 0.8$ \\
\textit{Type changed:}\\
SN2009jf & Ib & changed to Ic, see \citet{Gal-Yam2017}\\
SN2010al & IIn & changed to Ibn \citep{10al}\\
SN2008fq & IIn & changed to II\\
SN2007pk & IIn & changed to II\\
\tableline
\end{tabular}
\end{table*}

 \begin{table}
\centering
\caption{The SNe used for training in this work in addition to those from V19.\label{tab:mydata}}
\begin{tabular}{lllllll}
\tableline
       Name &         z &   Type &                        Photometry & \# Spec. & $E(B-V)_\mathrm{host}$ &                         Ref. \\
\tableline
   SN2013fs &  0.011855 &     II &        $UVW2,UVM2,UVW1,U,B,V,R,I$ &       23 &                  0.015 &                        (1-3) \\
  SN2017gpn &  0.007388 &    IIb &                       $B,V,g,r,i$ &        8 &                  0.000 &                          (4) \\
   SN2006el &  0.017000 &    IIb &                       $B,V,R,r,i$ &        6 &                  0.081 &                        (5-9) \\
   SN1996cb &  0.002372 &    IIb &                           $B,V,R$ &        8 &                  0.120 &                (5, 7, 10-11) \\
   SN2015ap &  0.011400 &    IIb &                     $B,V,u,g,r,i$ &       17 &                  0.000 &                        (4-5) \\
   SN2010jl &  0.010700 &    IIn &      $UVW2,UVM2,UVW1,U,B,V,u,r,i$ &        3 &                  0.022 &                      (12-16) \\
   SN2011fe &  0.000804 &     Ia &      $UVW2,UVM2,UVW1,U,B,V,R,I,r$ &       27 &                  0.014 &                  (13, 17-28) \\
   SN2012fr &  0.004000 &     Ia &  $UVW2,UVM2,UVW1,U,B,V,R,I,g,r,i$ &       22 &                  0.000 &          (13, 22, 24, 29-31) \\
  SN2017erp &  0.006174 &     Ia &        $UVW2,UVM2,UVW1,U,B,V,R,I$ &       11 &                  0.097 &         (13, 20, 22, 26, 32) \\
   SN2005cf &  0.006461 &     Ia &        $UVW2,UVM2,UVW1,U,B,V,R,I$ &       23 &                  0.100 &              (13, 26, 33-36) \\
   SN2012ht &  0.004000 &     Ia &  $UVW2,UVM2,UVW1,U,B,V,R,I,g,r,i$ &       14 &                  0.000 &              (13, 22, 37-39) \\
   SN2011by &  0.002843 &     Ia &        $UVW2,UVM2,UVW1,U,B,V,R,I$ &       11 &                  0.039 &      (13, 17, 20, 26, 40-41) \\
   SN2009ig &  0.008770 &     Ia &        $UVW2,UVM2,UVW1,U,B,V,R,I$ &       12 &                  0.000 &  (13, 17, 20, 22, 26, 42-43) \\
   SN2007af &  0.005464 &     Ia &        $UVW2,UVM2,UVW1,U,B,V,R,I$ &       14 &                  0.130 &          (13, 26, 34, 44-47) \\
   SN2004dk &  0.005200 &     Ib &                         $B,V,R,I$ &        8 &                  0.180 &                 (7-8, 48-50) \\
   SN2015ah &  0.016000 &     Ib &                   $B,V,u,g,r,i,z$ &        9 &                  0.020 &                        (4-5) \\
  SN2016jdw &  0.018900 &     Ib &                   $u,g,r,i,z,c,o$ &        5 &                  0.000 &                          (4) \\
  SN2017bgu &  0.008000 &     Ib &                   $u,g,r,i,z,c,o$ &        7 &                  0.020 &                          (4) \\
   SN1999ex &  0.011414 &    Ibc &                       $U,B,V,R,I$ &        5 &                  0.028 &                      (51-52) \\
   SN2006lc &  0.016200 &    Ibc &                     $B,V,g,r,i,z$ &        6 &                  0.298 &             (5-7, 13, 53-56) \\
  SN2019deh &  0.054690 &    Ibn &                         $g,r,i,z$ &        9 &                  0.224 &                      (9, 57) \\
 SN2019aajs &  0.035800 &    Ibn &                         $g,r,i,z$ &        4 &                  0.224 &                      (9, 57) \\
  SN2019iep &  0.057000 &    Ibn &                             $g,r$ &        7 &                  0.224 &                      (9, 57) \\
    SN2014L &  0.008029 &     Ic &                       $U,B,V,R,I$ &       13 &                  0.630 &                      (5, 58) \\
  SN2016coi &  0.003600 &  Ic-BL &  $UVW2,UVM2,UVW1,U,B,V,R,I,g,r,i$ &       48 &                  0.000 &                  (13, 59-61) \\
   SN2003jd &  0.018860 &  Ic-BL &                         $B,V,R,I$ &       10 &                  0.100 &                    (5-7, 62) \\
\tableline
\end{tabular}

\tablerefs{(1) \cite{2016MNRAS.459.3939V}; (2) \cite{2017NatPh..13..510Y}; (3) \cite{2016PASA...33...55C}; (4) \cite{2019MNRAS.485.1559P}; (5) \cite{2019MNRAS.482.1545S}; (6) \cite{2014ApJS..213...19B}; (7) \cite{2014AJ....147...99M}; (8) \cite{2011ApJ...741...97D}; (9) median $E(B-V)_\mathrm{host}$ value, \cite{2016MNRAS.458.2973P}; (10) \cite{1999AJ....117..736Q}; (11) \cite{2001AJ....121.1648M}; (12) \cite{2012AJ....144..131Z}; (13) \cite{2014Ap&SS.354...89B}; (14) \cite{2016MNRAS.456.2622J}; (15) \cite{2014ApJ...797..118F}; (16) \cite{2012AJ....143...17S}; (17) \cite{2016PASP...128...961}; (18) \cite{2013A&A...554A..27P}; (19) \cite{2016ApJ...820...67Z}; (20) \cite{2019MNRAS.490.3882S}; (21) \cite{2013NewA...20...30M}; (22) \cite{2020MNRAS.492.4325S}; (23) \cite{2012ApJ...752L..26P}; (24) \cite{2017MNRAS.472.3437G}; (25) \cite{2014MNRAS.439.1959M}; (26) \cite{2012MNRAS.425.1789S}; (27) \cite{2011Natur.480..344N}; (28) \cite{2015MNRAS.446.3895F}; (29) \cite{2013ApJ...770...29C}; (30) \cite{2018arXiv180310095C}; (31) \cite{2014AJ....148....1Z}; (32) \cite{2019ApJ...877..152B}; (33) \cite{2009ApJ...697..380W}; (34) \cite{2010ApJS..190..418G}; (35) \cite{2007A&A...471..527G}; (36) \cite{2007MNRAS.376.1301P}; (37) \cite{2018PASP..130f4101V}; (38) \cite{2014ApJ...782L..35Y}; (39) \cite{2018arXiv180906381B}; (40) \cite{2013MNRAS.430.1030S}; (41) \cite{2020MNRAS.491.5991F}; (42) \cite{2012ApJ...744...38F}; (43) \cite{2012ApJ...749...18B}; (44) \cite{2007ApJ...671L..25S}; (45) \cite{2011Sci...333..856S}; (46) \cite{2010ApJ...721.1608B}; (47) \cite{2012AJ....143..126B}; (48) \cite{2017PASP..129e4201S}; (49) \cite{2008ApJ...687L...9M}; (50) \cite{2008A&A...488..383H}; (51) \cite{2002AJ....124..417H}; (52) \cite{2002AJ....124.2100S}; (53) \cite{2017arXiv170707616S}; (54) \cite{2011A&A...526A..28O}; (55) \cite{2014arXiv1401.3317S}; (56) \cite{2018A&A...609A.135S}; (57) Kool et al. in preparation; (58) \cite{2018ApJ...863..109Z}; (59) \cite{2019ApJ...883..147T}; (60) \cite{2018MNRAS.478.4162P}; (61) \cite{2018MNRAS.473.3776K}; (62) \cite{2008MNRAS.383.1485V}}
\end{table}

 \begin{table}
\centering
\caption{All SNe used for training in this work and the DQI calculated for them.\label{tab:alldata}}
\begin{tabular}{lp{15cm}}
\tableline
  Type &                                                                                                                                                                                                                                                                                                                                                                           Names (DQI) \\
\tableline
    II &  ASASSN14jb (0.91), ASASSN15oz (0.89), SN1999em (0.98), SN2004et (0.83), SN2005cs (0.96), SN2007od (0.91), SN2007pk (0.92), SN2008bj (0.83), SN2008fq (0.95), SN2009N (0.84), SN2009bw (0.83), SN2009ib (0.87), SN2009kr (0.90), SN2012A (0.97), SN2012aw (1.00), SN2013ab (0.99), SN2013am (0.97), SN2013by (0.95), SN2013ej (1.00), SN2013fs (1.00), SN2014G (1.00), SN2016X (1.00) \\
   IIb &                                                                                                                                          SN1993J (1.00), SN1996cb (0.87), SN2006T (0.95), SN2006el (0.84), SN2008ax (1.00), SN2008bo (0.91), SN2011dh (1.00), SN2011ei (0.95), SN2011fu (1.00), SN2011hs (0.96), SN2013df (0.85), SN2015ap (0.99), SN2016gkg (0.95), SN2017gpn (0.97) \\
   IIn &                                                                                                                                                                                                                                                                                                                                                      SN2006aa (0.92), SN2010jl (0.90) \\
    Ia &                                                                                                                                                                                                                                               SN2005cf (1.00), SN2007af (0.93), SN2009ig (0.91), SN2011by (0.99), SN2011fe (0.99), SN2012fr (1.00), SN2012ht (0.99), SN2017erp (0.99) \\
    Ib &                                                                                                                                                         SN1999dn (0.90), SN2004dk (0.87), SN2004gq (0.92), SN2004gv (0.90), SN2005hg (0.95), SN2006ep (0.86), SN2007Y (0.96), SN2007uy (0.83), SN2009iz (0.91), SN2015ah (0.92), SN2016jdw (0.90), SN2017bgu (0.86), iPTF13bvn (1.00) \\
   Ibc &                                                                                                                                                                                                                                                                                                                    SN1999ex (0.91), SN2004gt (0.80), SN2006lc (0.87), SN2012au (0.81) \\
   Ibn &                                                                                                                                                                                                                                                                                                                SN2010al (0.94), SN2019aajs (0.88), SN2019deh (0.98), SN2019iep (0.97) \\
    Ic &                                                                                                                                                                                                                                                                   SN1994I (0.96), SN2004aw (0.85), SN2004fe (0.96), SN2007gr (0.96), SN2009jf (0.94), SN2013ge (0.99), SN2014L (0.94) \\
 Ic-BL &                                                                                                                                                                                                                                               SN1998bw (0.89), SN2002ap (0.99), SN2003jd (0.89), SN2006aj (0.93), SN2007ru (0.88), SN2009bb (0.92), SN2012ap (0.91), SN2016coi (0.99) \\
\tableline
\end{tabular}
\end{table} 
For the SNe we added, spectra were taken from WISeREP\footnote{\url{https://wiserep.weizmann.ac.il}} \citep{wiserep} and photometry from the Open Supernova Catalog\footnote{\url{https://sne.space}} \citep{osc}, where available. The values for Milky Way reddening $E(B-V)_\mathrm{MW}$ were taken from \citet{ebvmw}. The host reddening estimations $E(B-V)_\mathrm{host}$ were taken from the provided references. When no estimation could be found, the median value for the SN type was taken.

\subsection{Data Quality Index}
The quality of the output of \texttt{PyCoCo} depends on the coverage of the spectral-temporal region $T\times\Lambda$ by the spectroscopic and photometric data of each object. A Data Quality Index (DQI) between 0 and 1 is thus calculated for each SN in the following manner. For each point on the $T\times\Lambda$ grid, the distance to the nearest input data point (spectroscopic or photometric) is calculated as the usual Euclidean distance, where the time and wavelength axes are normalized by their total length. The fraction of grid points which are closer than 0.1 to a data point is defined as the DQI. We only included SNe with ${\rm DQI} \geq 0.8$ in our analysis. The impact of this choice is minimal (the removal of 4 objects; Table~\ref{tab:vincmod}).

\section{Results}

\begin{figure*}
\myplot{1.1}{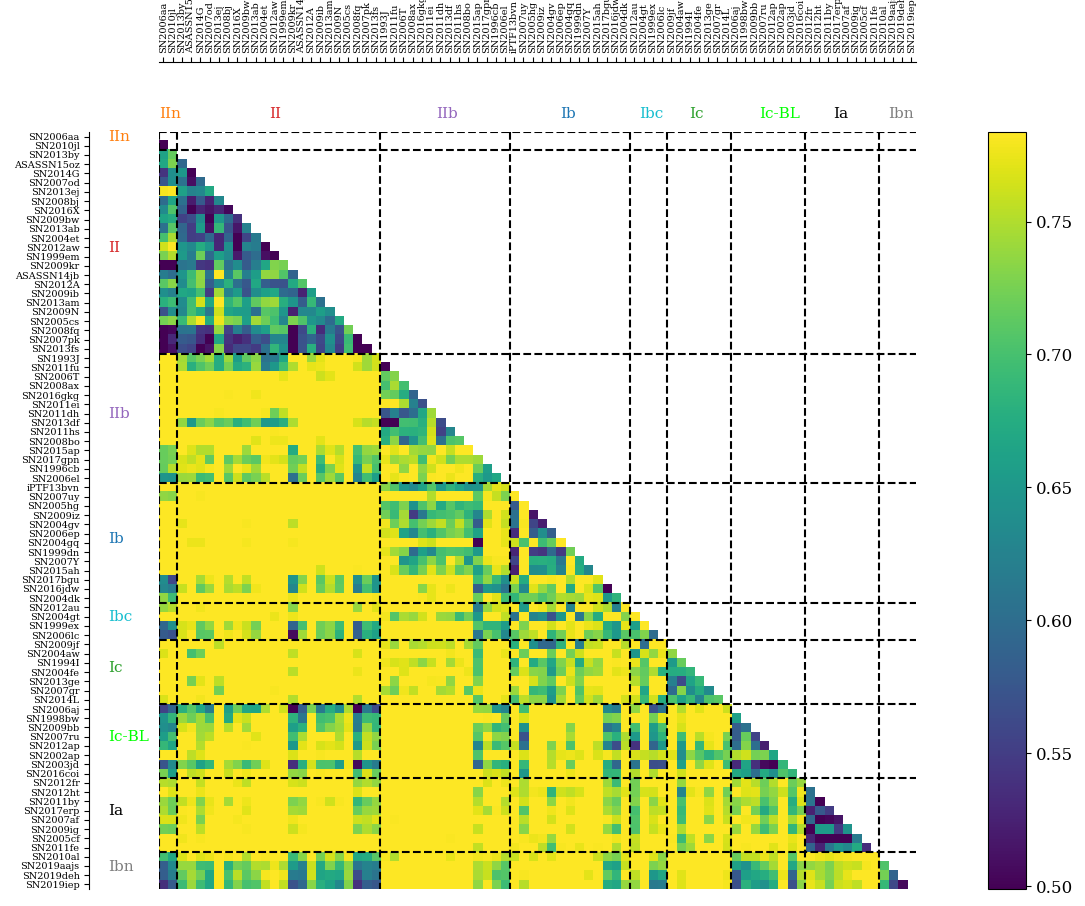}
    \caption{The resulting dissimilarity matrix after applying the RF dissimilarity measure on the data set. Colors represent dissimilarity, are in linear scale and only vary between the 1st and 50th percentile.}
    \label{fig:dissim}
\end{figure*}

After applying the EMPCA algorithm, the fraction of data variance explained is $93\%$. Figure~\ref{fig:dissim} shows the distance matrix $D$ resulting from an unsupervised RF as described. It is not completely symmetrical due to computation errors -- the median absolute symmetric difference of off-diagonal terms is $0.005$ -- so we enforce symmetry by using $\frac{1}{2}(D+D^\mathrm{T})$ as the dissimilarity matrix. A certain degree of block-diagonality is noticeable, expressing agreement with the current classification scheme for these well-documented SNe.

\subsection{Visualization of The Metric Space}\label{sec:vis}

\begin{figure*}
    \makebox[\textwidth][c]{\includegraphics[trim=30 20 0 40, clip]{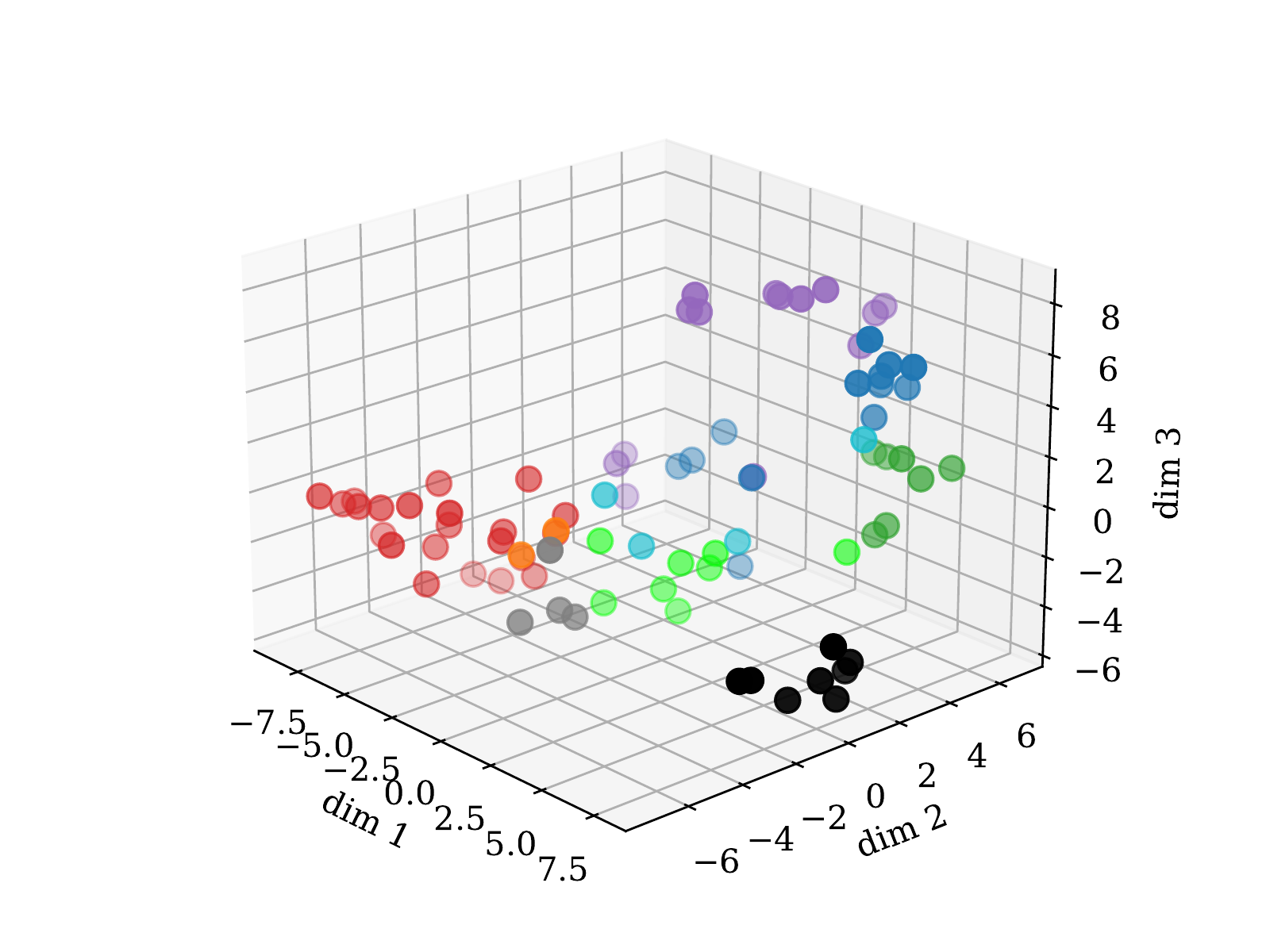}}
    \myplot{0.8}{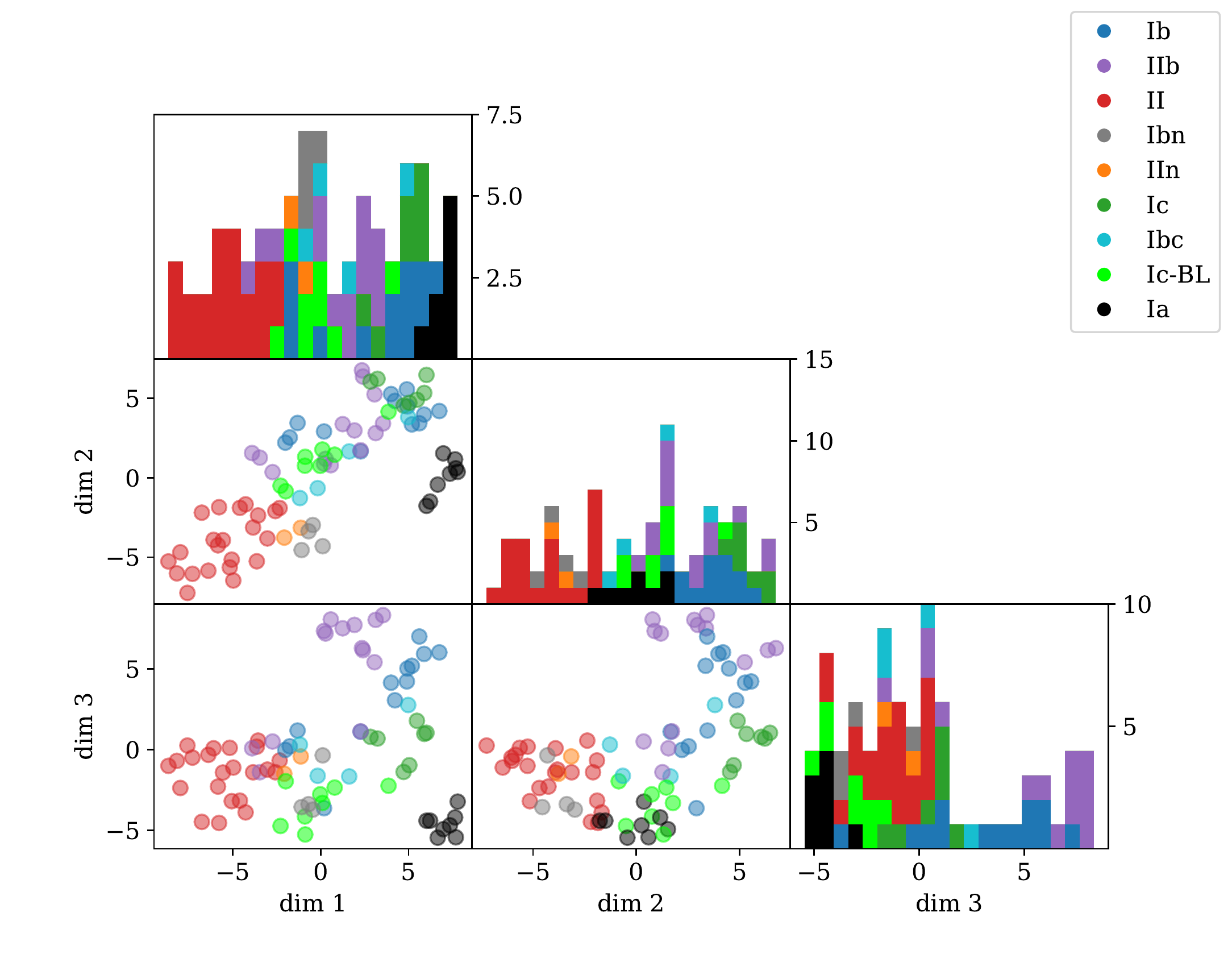}
    \caption{Embedding of the dissimilarity matrix in Figure~\ref{fig:dissim} by tSNE with parameters \texttt{perplexity = 10, learning\_rate = 10}, shown in 3D (top) and in a corner plot (bottom). The KL divergence with respect to the dissimilarity matrix is $0.20$. The colors represent the types as assigned for the spectral series by a human expert. The axes represent the coordinates of the embedded Euclidean space. A rotatable version of the top panel is available in the online version and under \url{https://github.com/ofek-b/spectra_in_time/blob/rf_dlog_no_pca/3d_paper.html}. The rotation is about the dim 3 axis. The speed and direction of the rotation can be controlled by the buttons or by dragging the slider.}
    \label{fig:tsne_3d}
\end{figure*}

For visualisation of this dissimilarity matrix inside a low-dimensional Euclidean space while trying to preserve structure, we use t-Distributed Stochastic Neighbor Embedding (tSNE), developed by \citet{tsne}. The visualization in 3d space is shown in Figure~\ref{fig:tsne_3d}, with colors representing the spectroscopic type assigned by a human expert after manually inspecting the evolution of spectra available for the SN. Examining the visualization one can notice the aforementioned block-diagonality in the form of well-separated clusters which are relatively homogeneous in type. We remind the reader that the spectroscopic types represented by the colors were not used in the analysis which led to the dissimilarity matrix or the visualization (the only exception is mentioned in Section~\ref{subsec:pycoco}).

\begin{figure*}
\myplot{1}{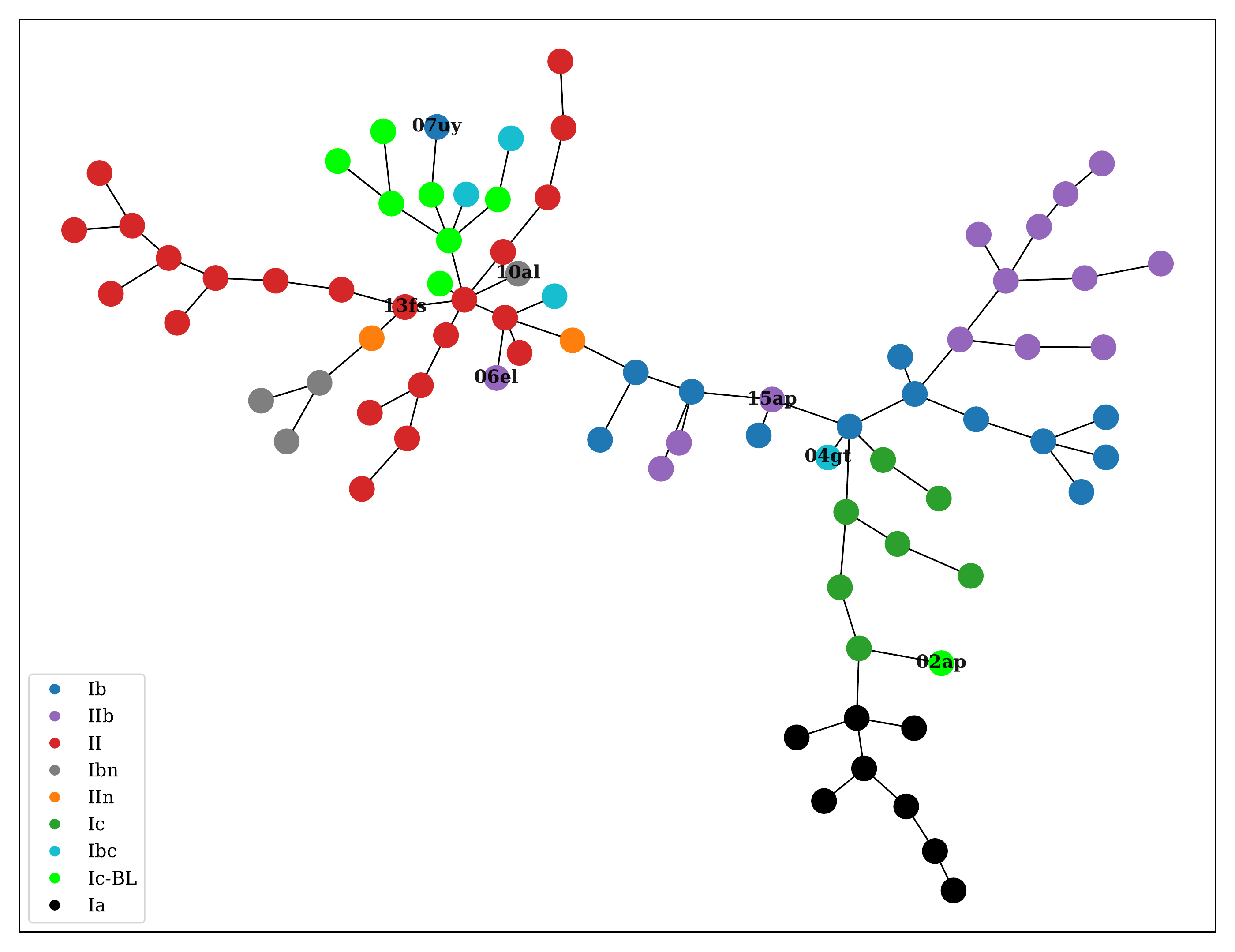}
    \caption{The MST of the graph of SNe defined by the dissimilarity matrix. Note that edge lengths are arbitrary.}
    \label{fig:mst}
\end{figure*}

Another way to visualize the results is to look at the Minimum Spanning Tree (MST) of the fully-connected graph defined by the distance matrix. The MST is a fully-connected subgraph minimal in the sum of edge weights\footnote{The MST is generally not unique, though it is unique in this case.} (here -- distances) and is shown in Figure~\ref{fig:mst}. The algorithm by \citet{yamada}\footnote{\url{https://github.com/dakota-hawkins/yamada}} was used. The concept of MST in the context of an abstract dissimilarity space could be employed, as was done by \citet{sequencer}, to extract sequences from the data, which could later be compared with physical parameters.

Looking at Figure~\ref{fig:tsne_3d}, we see several remarkable features. Objects of Type Ibn are close to those of Type IIn probably due to sharing narrow features. In contrast, the proximity of Type Ic-BL objects to the Type II cluster is unexpected. A possible explanation could be that the features those two groups share with one another are the dominance of the continuum.

One notices the very close pair of objects SN2015ap (IIb, purple) and SN2004gq (Ib, blue). A possible explanation is that SN2015ap, although exhibiting H-absorption initially, loses them at a relatively early stage, $\sim 10$ days after explosion, leaving no H evident in its spectrum. Another object to be noticed is SN2004gt (Ibc, cyan) in between the main Ib and Ic clusters, which fits previous classification \citep{04gt_ibc}. As for SN2007Y (main Ib cluster, blue), its initial classification was Ib despite H$\alpha$ absorption lines observed at early times and it has been claimed that IIb would be a better classification \citep{late_IIb,07Y,IIb_07Y}. The embedding suggests that it is part of the main Ib cluster, though it is on its edge.

As could also be observed in the tSNE visualization, the Type~Ic SN2009jf and SN2007gr (two dark green points in the upper main Ic cluster) are very close by, which fits the fact that they are almost identical spectroscopically in the optical and IR wavelengths throughout their evolution \citep{09jf_07gr}. They are also close to the Ib cluster, with the He-poor SN2007gr slightly differing from SN2009jf which shows stronger He lines, leading \citet{09jf_07gr} to classify as Ib, though they acknowledge the possibility that a Ic could be a better choice. Likewise, \citet{on_Ibc} have found that the data of SN2007gr, classified there as Type Ic, may be well explained by a He-rich model, but with weak lines. These authors conclude from their analysis of Type Ib and Ic SNe that the property usually assumed to define Type Ic event, the lack of He, is justified for very few, if any, objects. Indeed, a continuum such as the one visualized in Figure~\ref{fig:tsne_3d}, may be a more suitable way to understand those objects.

\subsection{The Split of Type~Ib SNe}

\begin{figure*}
    \myplot{1}{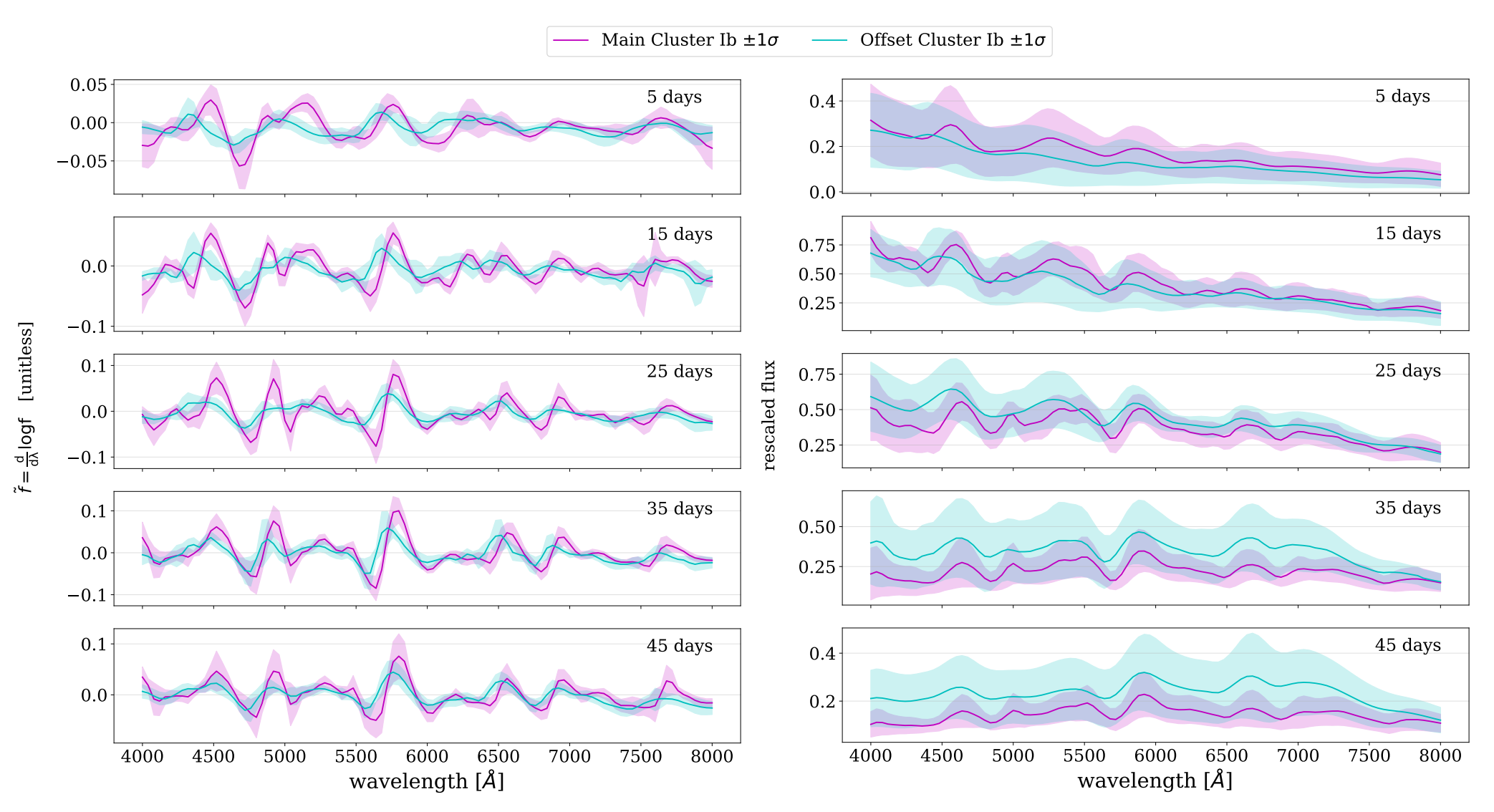}
    \caption{Comparison between the mean spectra as a function of time from explosion of Type~Ib SNe from the main cluster and of the rest of the events (see text) in the offset sub-cluster. The flux $f$ (right panel) and its derivative $\tilde{f}$ (left panel) are used for the comparison. One can notice (more clearly on the left hand side) that the spectral features are narrower, more prominent and have lower expansion velocity, especially before and at peak, for SNe that belong to the main cluster than for those from the offset one.}
    \label{fig:Ib_split}
\end{figure*}

It can be noticed in the visualizations that some Type~Ib SNe are set apart from the main cluster formed by this type. The SNe in the main cluster are SN2004gv, SN2007Y, SN2006ep, SN2015ah, SN1999dn, iPTF13bvn, SN2005hg and SN2009iz. The offset Type~Ib SNe are SN2004gq, SN2007uy, SN2016jdw, SN2017bgu, SN2004dk and SN2015ap, which is of Type~IIb but still included in this list (see Section~\ref{sec:vis}). Figure~\ref{fig:Ib_split} shows a comparison between the mean spectra of these two groups.

The figure suggests that this split could be correlated with the SN expansion velocity, as the features in the mean spectrum of the offset cluster are wider and bluer, especially before- and at peak. To validate this effect, we examine the raw spectra of each event near peak (i.e., prior to any processing of the data). We find that the average expansion velocities derived from the location of the absorption minimum of the He I lines at $\lambda\lambda5876,6678,7065$\AA\ are $9700\pm1000$ and $12300\pm1200$\,km\,s$^{-1}$ for the main and offset clusters respectively. The velocities were measured using the dedicated tool on WISeREP. This supports the results of our analysis, and suggests that there may be a split between populations of low-velocity and high-velocity SNe Ib. This of course brings to mind the well-known separation between spectroscopically normal SNe Ic and broad-line, high-velocity SNe Ic-BL (e.g.,  \citealt{Modjaz2016}, \citealt{Gal-Yam2017}, \citealt{phys_mot}), though for SNe Ib the velocity spread between normal and high-velocity events may be less extreme, making this division less obvious.

\subsection{An Application to SNe With Less Data} \label{sec:displ}
\begin{figure*}
    \plottwo{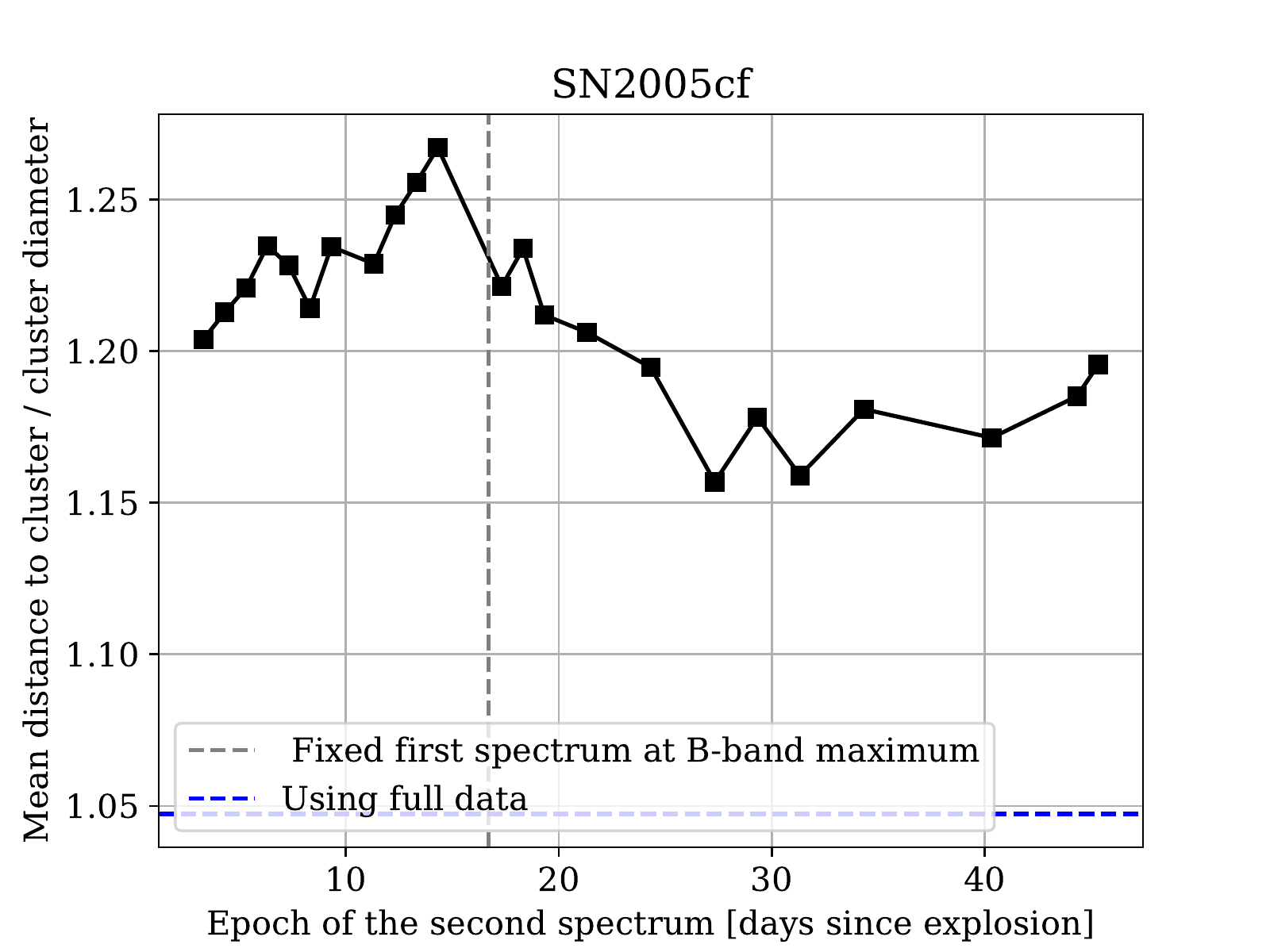}{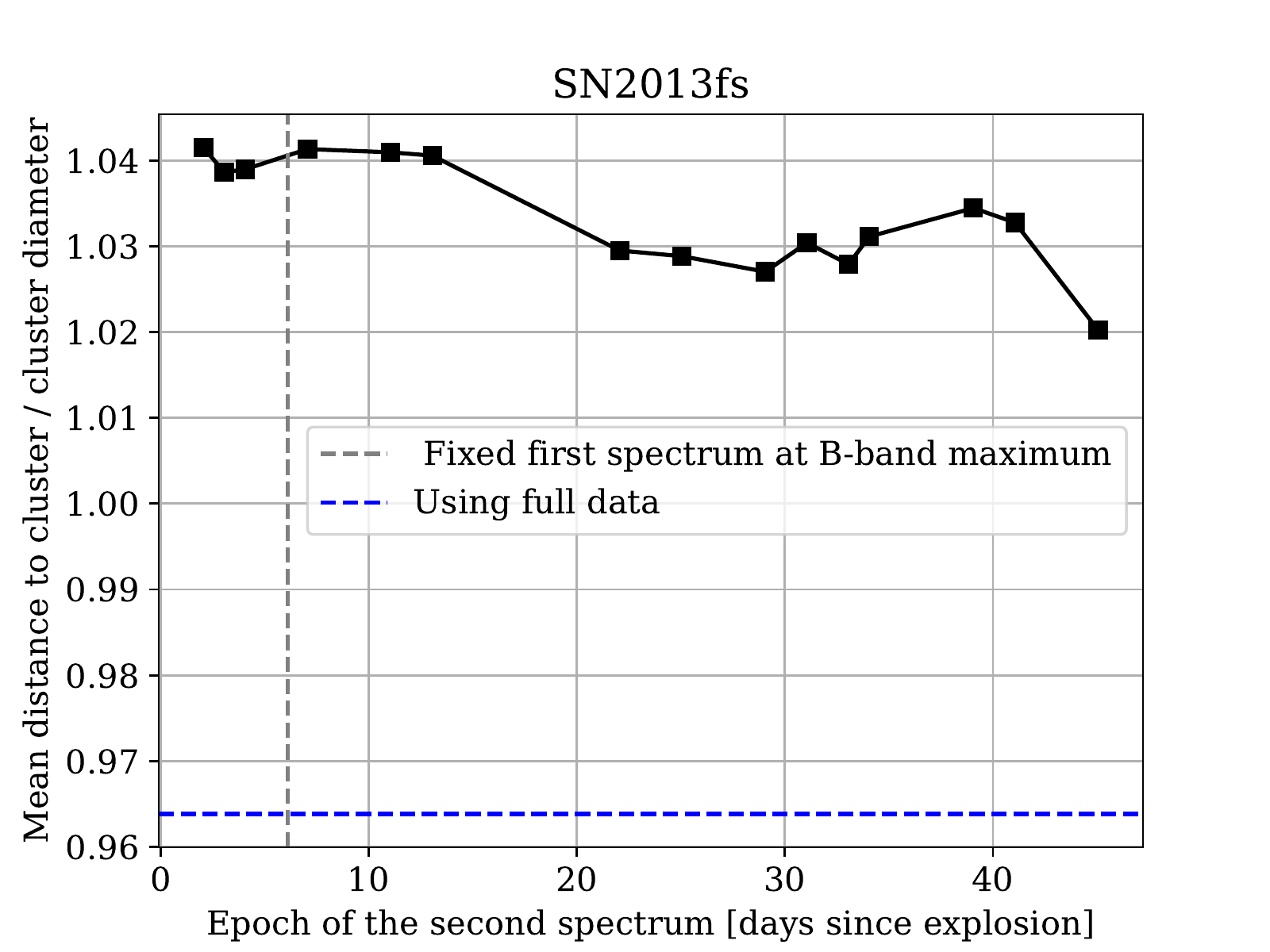}
    \caption{Embedding SN2005cf (Type Ia) and SN2013fs (Type II) with degraded data in a metric space trained on a set not containing these SNe. After training, each SN was embedded in the space using only a subset of the data. The degraded subsets are the  \texttt{PyCoCo} output for the entire photometric data, but only two spectra: one at the B-band maximum and the other at the time given on the x-axis. The y-axis is the distance of the embedded degraded form from the respective cluster (Type~Ia or Type~II SNe), in units of the specific cluster diameter. One can see that the degraded data sets remain in the close vicinity of the mother cluster.}
    \label{fig:degraded}
\end{figure*}

While the SNe used for training are some of the best observed objects of their types, we wish to test to which extent is the embedding of a new SN with lower amount of data useful. Figure~\ref{fig:degraded} shows the effect of only including two spectra in the input to \texttt{PyCoCo} on the displacement of the embedding relative to the "true" one produced by using all of the data for the SN. We test this on the Type~Ia SN2005cf and the Type~II SN2013fs. We do not degrade the photometric data in this test. It can be seen that at least for those SNe, using only two spectra does not affect much the quality of embedding. This is highly relevant since future data sets from ongoing and future surveys such as the Zwicky Transient Facility (ZTF; \citealt{Bellm2019, Graham2019}) and in the future the Rubin Observatory include high-quality multicolor light curves. Spectroscopy is often obtained as soon as possible after discovery (e.g., \citealt{Gal-Yam2011}) and again around peak, so a typical future data set may likely include this combination of two spectra and good light curves.

\section{Conclusion}

We present a method for quantitative comparison of SNe by means of an unsupervised embedding in a metric space. This data-driven method identifies the important correlations/spectral features evident in the spectral-temporal energy distributions it is trained on. Other methods which have been used for quantitative classification of SNe \citep[e.g.,][]{sun2017,phys_mot} can be understood as supervised decision trees constructed by experienced astrophysicists rather than by an automated training process. Our approach is in principal similar, but chooses an unsupervised method in order to detect the most important correlations -- deemed as such by an objective measure -- in a given training set. The method succeeds in defining some clusters, which mostly agree with the known spectroscopic Types, and in visualizing the IIb-Ib-Ic-Ic-BL continuum. There, objects like SN2007Y, SN2009jf and SN2015ap could get a quantitative answer regarding their Type by this method, while the classical element-based classification is not as conclusive.

One result our method produces is the possible split of Type Ib SNe into two subgroups, of normal and higher-velocity events, perhaps mimicking the more prominent division between spectroscopically normal SNe Ic and SNe Ic-BL. As discussed earlier, this separation could be of a discrete nature and should be investigated further (preferably with a larger sample) to further validate or dismiss its existence and test the underlying differences in the physics of the explosions.

Preliminary tests show that even a small number of spectra, when combined with multi-band photometric measurements, suffice for obtaining a relatively accurate embedding. This could make the method useful in the context of surveys such as those to be conducted by the Rubin Observatory.

The application as a classifier for less typical SNe should be approached with caution, as the training set needs to contain a wide enough selection of SNe (which in turn need to have enough data) in order to be able to produce a meaningful result.

\begin{acknowledgments}
\section*{Acknowledgments}
We thank Dalya Baron, Ido Irani, Barak Zackay and the ZTF collaboration for useful advice, and an anonymous referee for a helpful review.
    
\software{This work made use of the Python packages \texttt{NumPy} \citep{numpy}, \texttt{Matplotlib} \citep{matplotlib}, \texttt{scikit-learn} \citep{scikit-learn}, \texttt{pandas} \citep{pandas} and \texttt{NetworkX} \citep{networkx}.}
\end{acknowledgments}

\bibliography{refs,mydata}

\end{document}